\documentclass[%h
aip,
jcp,
 sd,%
amsmath,amssymb,
reprint,%
]{revtex4-1}

\usepackage{graphicx}% Include figure files
\usepackage{dcolumn}% Align table columns on decimal point
\usepackage{bm}% bold math
\usepackage{todonotes}
\usepackage{color}

\newcommand{\new}{\nonumber\\}

\newcommand{\ave}[1]{\left\langle #1 \right\rangle}

\usepackage{amsmath}	% required for `\align' (yatex added)
\begin{document}

\preprint{APS/123-Qed} \title{Note: Effect of localization on mean-field
 density of state near jamming}

\author{Harukuni Ikeda}
 \email{harukuni.ikeda@lpt.ens.fr}
\affiliation{%
\'Ecole Normale Sup\'erieure, UMR 8549 CNRS, 24 Rue Lhomond, 75005 Paris, France
}%
% \author{Pierfrancesco Urbani}
% \affiliation{%
% Institut de physique th\'eorique, Universit\'e Paris Saclay,
%  CNRS, CEA, F-91191 Gif-sur-Yvette, France
% }%
% \author{Francesco Zamponi}
% \affiliation{%
% \'Ecole Normale Sup\'erieure, UMR 8549 CNRS, 24 Rue Lhomond, 75005 Paris, France
% }%

\date{\today}% It is always \today, today,

\pacs{05.20.-y, 61.43.Fs, 63.20.Pw}% PACS, the Physics and Astronomy
%\pacs{64.70.Q-, 05.20.-y, 64.70.Pf}% PACS, the Physics and Astronomy

\newcommand{\PP}{\mathcal{P}}

\maketitle

%\section{Known result}
The perceptron is a mean-field model of the jamming
transition~\cite{franz2017universality}.  The model is simple enough to
analytically determine the critical exponents of several physical
quantities such as the contact number and gap distributions functions.
The predicted exponents are the same of those of the hard spheres in the
large dimension limit and also reasonably close to the numerical results
in finite dimensions~\cite{charbonneau2014fractal}. The simplicity of
the model also allows us to analytically calculate the density of state
$D(\omega)$, which is the distribution of the eigenvalue spectrum of the
Hessian matrix~\cite{franz2015universal}. Near the jamming point, the
model predicts for $\omega\ll 1$~\footnote{We omitted proportional
constants and high frequency cutoff, which are not relavant to the
present discussions.}
\begin{align}
 D(\omega) \sim \frac{\omega\sqrt{\omega^2-\omega_{\rm min}^2}}{\omega^2
 + \omega_*^2}\theta(\omega-\omega_{\rm min}).\label{164344_2Aug18}
\end{align}
where $\theta(x)$ is the Heaviside step function and
\begin{align}
 &\omega_*^2 = c_1 \delta z^2,\new
 &\omega_{\rm min}^2 = c_2 \delta z^2 -c_3 p.\label{161255_2Aug18}
\end{align}
$p$ is the pressure and $\delta z=z-z_{\rm iso}$ is the deviation of the
contact number $z$ from the isostatic value $z_{\rm iso}$. $c_1$, $c_2$
and $c_3$ are constants. % The perceptron model and also variational
% argument predict $\delta z \sim
% p^{1/2}$~\cite{wyart2005effects,yan2016variational,franz2017universality}.
Essentially the same result of eq.~(\ref{164344_2Aug18}) is also
obtained by the effective medium theory, except the trivial Debye
modes~\cite{degiuli2014effects}.  The mean-field perceptron model
predicts $\omega_{\rm min}=0$ sufficiently near the jamming point, which
means $p=c_2\delta z^2/c_3$. In this case, the scaling behavior of
$D(\omega)$ near the jamming point is
\begin{align}
D(\omega)\sim
\begin{cases}
 {\rm constant} & (\omega \gtrsim \omega_*)\\
 (\omega/\omega_*)^2 & (\omega\ll \omega_*).
\end{cases} 
\label{235158_1Aug18} 
\end{align}
However, it has been revealed that the mean-field prediction of
$D(\omega)$ is inconsistent with the numerical result in finite
dimensions~\cite{PhysRevLett.117.035501,mizuno2017continuum,shimada2018spatial}.
Recent numerical studies in finite dimensions show that, if one
carefully removes the phonon mode that follows the Debye low
$\omega^{d-1}$, one obtains~\cite{mizuno2017continuum}
\begin{align}
 D(\omega)\sim
 \begin{cases}
 {\rm constant} & (\omega \gtrsim \omega_*)\\
 (\omega/\omega_*)^2 & (\omega_{\rm ex0}\ll \omega \ll \omega_*),\\
  (\omega/\omega_*)^4 & (\omega\ll\omega_{\rm ex0}),
\end{cases}\label{161302_2Aug18} 
\end{align}
where $\omega_{\rm ex0}\sim \delta z$ but the proportional constant is
much smaller than that of $\omega_*$. In this note, relying on a bit
empirical argument, we reconcile the above discrepancy between the
mean-field and finite dimensional results.

The reason of the discrepancy between the mean-field and finite
dimensional results is twofold.  (i) In finite dimensions, the system is
not exactly marginally stable and the pre-stress is smaller than that
required by the marginal stability, $p<p_* \equiv c_2\delta
z^2/c_3$~\cite{degiuli2014effects,lerner2014breakdown}. We introduce the
distance to the marginal stability as
\begin{align}
 \varepsilon \equiv (p_*-p)/p_*.
\end{align}
(ii) In finite dimensions, the eivenvectors for $\omega<\omega_{\rm
ex0}$ are localized in space~~\cite{mizuno2017continuum}, not as in case
of the mean-field model where all the modes are extended. This allows us
to separate the system into several parts and each of them may have a
different value of $\varepsilon$. To express this fluctuation, we borrow
a rather old idea by Gurevich \textit{et al.}.~\cite{PhysRevB.67.094203}
and more recently Ji \textit{et al.}~\cite{ji2018theory}, where they
modeled the localized modes by the anharmonic oscillators with
different frequencies and $D(\omega)$ is calculated by summing up the
contributions of them. Interestingly, with proper assumptions, this
approach correctly reproduces the $\omega^4$ scaling for the small value
of $\omega$, though it is not clear how to apply it to the jamming
transition. As in case of Ji \textit{et al.}~\cite{ji2018theory}, we
consider the distribution function of $\varepsilon$, $\PP(\varepsilon)$,
which is normalized so that $\int_0^\infty d\varepsilon
\PP(\varepsilon)=1$.  We set the small cutoff $\varepsilon_{\rm ex0}\ll
1$ and assume that $\PP(\varepsilon) =O(\varepsilon_{\rm ex0}^{-1})$ for
$\varepsilon\lesssim\varepsilon_{\rm ex0}$ and $\PP(\varepsilon)\sim 0$ for
$\varepsilon \gg \varepsilon_{\rm ex0}$.  Then, the mean value of the
density of state is calculated as $D(\omega)=\int_{0}^\infty
d\varepsilon \PP(\varepsilon) D(\omega,\varepsilon)$.  Below, using the
above assumptions, we show that the scaling behavior of $D(\omega)$,
eq.~(\ref{161302_2Aug18}), is correctly reproduced including the scaling
factors $\omega_*$.

\begin{figure}[t]
\includegraphics[width=8cm]{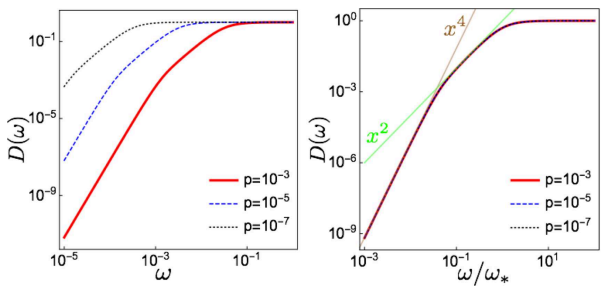} \caption{The density of state
 $D(\omega)$: (left) 
 The results for $p=10^{-3}$, $10^{-5}$ and $10^{-7}$. (right) The
 scaling plot for the same data. The green and brown lines denote the
 scaling behaviors expected from the asymptotic analysis, see main
 text.}  \label{171248_2Aug18}
\end{figure}

We first discuss the scaling behavior in the low frequency limit,
 $\omega^2/p_* \ll \varepsilon_{\rm ex0}$. Substituting
 $p=(1-\varepsilon)p_*$ into eq.~(\ref{164344_2Aug18}), and averaging
 over $\varepsilon$, we obtain
\begin{align}
D(\omega) 
 %& \equiv 
 % \int_0^\infty d\varepsilon \PP(\varepsilon)D(\omega,\varepsilon) \new
 &\sim \int_0^\infty d\varepsilon \PP(\varepsilon)
\frac{\omega\sqrt{\omega^2-c_3\varepsilon p_*}}{\omega^2
 + \omega_*^2}\theta(\omega-\sqrt{c_3\varepsilon p_*}) \new 
 &\sim \omega_*^{-2}\PP(0)\int_0^{\omega^2/(c_3p_*)} d\varepsilon 
 \omega\sqrt{\omega^2-c_3 \varepsilon p_*}\new
 &\sim \varepsilon_{\rm ex0}^{-1}(\omega/\omega_*)^4.\label{173412_2Aug18}
\end{align}
Defining $\omega_{\rm ex0} \equiv \sqrt{\varepsilon_{\rm ex0}p_*}$, one
can see that the above scaling is the same of that of the
$\omega\ll\omega_{\rm ex0}$ regime of eq.~(\ref{161302_2Aug18}). With
the similar calculations, one can confirm that the scaling for
$\omega\gg \omega_{\rm ex0}$ is unchanged from the mean-field result,
eq.~(\ref{235158_1Aug18}). Thus, we recovered the same scaling behaviors
of eq.~(\ref{161302_2Aug18}). Finally, for concreteness, in
Fig.~\ref{171248_2Aug18}, we show the numerical result of $D(\omega)$
obtained by assuming $\PP(\varepsilon)= \varepsilon_{\rm
ex0}^{-1}e^{-\varepsilon/\varepsilon_{\rm ex0}}$, where
$\varepsilon_{\rm ex0}= 10^{-3}$ and $c_1=c_2=c_3=1$. If one rescales
$\omega$ by $\omega_*$, all the data are collapsed on a single curve as
expected from eq.~(\ref{161302_2Aug18}).

In summary, we discussed the effects of the localized modes on the
density of state $D(\omega)$ by considering the probability distribution
function of the proximity to the marginal stability
$P(\varepsilon)$. Our calculation reproduces the finite dimensional
numerical result near the jamming point,
eq.~(\ref{161302_2Aug18}). In particular, the theory successfully
captures the novel $D(\omega)\sim \omega^4$ scaling including its
pressure dependence of the pre-factor, see
eq.~(\ref{173412_2Aug18}). Note, the derivation of
eq.~(\ref{173412_2Aug18}) does not depend on the precise form of
$\PP(\varepsilon)$. If $\PP(\varepsilon)$ is a finite and continuous
function at $\varepsilon=0$, one always gets $D(\omega)\sim \omega^4$
for small $\omega$. This may explain the robustness of the $\omega^4$
scaling against the different interaction potentials and
dimensions~\cite{kapteijns2018universal,wang2018low,PhysRevLett.117.035501,mizuno2017continuum,shimada2018spatial,wang2018low}.

There are still several important points that deserve investigation.  A
tentative list is the following:
\begin{itemize}
 \item The cutoff $\varepsilon_{\rm ex0}$ is related to the average
       value of $\varepsilon$ as $ \varepsilon_{\rm ex0} \sim
       \int_0^\infty d\varepsilon\PP(\varepsilon)\varepsilon \equiv
       \ave{\varepsilon} $. It is reported that in the two dimensional
       packing near the jamming point, $\ave{\varepsilon}\approx 0.04$~\cite{degiuli2014effects,lerner2014breakdown}.  We expect
       $\ave{\varepsilon}$ decreases with increasing the dimension since
       the localized modes are suppressed in high
       dimensions~\cite{PhysRevLett.117.045503}. Its dimensional
       dependence deserves further investigation.

 \item The scaling of the lowest frequency is changed only if
       $\PP(\varepsilon)$ is \textit{not} finite at $\varepsilon=0$. For
       instance, when $\PP(\varepsilon) \sim A\varepsilon^{-\alpha}$ for
       small $\varepsilon$, eq.~(\ref{173412_2Aug18}) is replaced by
       $D(\omega)\sim A(\omega/\omega_*)^{4-2\alpha}$. This can
       correspond to the configuration obtained by quenching from very
       high temperature. In this case, Lerner and
       Bouchbinder~\cite{PhysRevE.96.020104} observe $D(\omega)\sim
       \omega^{\beta}$ with $\beta < 4$ suggesting that
       $\alpha>0$. However, a more recent numerical result shows that
       $\beta=4$ for wide range of the initial
       temperature~\cite{wang2018low}, the initial temperature just
       affects the pre-factor. Further numerical investigations are
       necessary to determine which of the two scenarios is correct.
       
 \item We assume that the system can be divided into several
       sub-components. Since our theory does not take into account the
       interactions between the sub-components, the typical length scale
       of them should be much larger than the correlation length of the
       system. However, there are many different lengths have been
       proposed for the jamming transition and it is not very clear
       which length scale would be relevant to the current argument.
       The size of the localized excitation recently investigated by
       Shimada \textit{et al.}~\cite{shimada2018spatial} might be a
       promising candidate.

 % \item We assume that $D(\omega)$ of the sub-components follows
 %       eq.~(\ref{164344_2Aug18}). This assumption is reasonable for high
 %       enough dimensions since eq.~(\ref{164344_2Aug18}) holds exactly
 %       in the large dimension limit.  However, in two or three
 %       dimensions, it is not clear whether $D(\omega)$ of the
 %       sub-component has the form of eq.~(\ref{164344_2Aug18}). In
 %       finite dimensions, a more widely accepted approach, which
 %       reproduces the $\omega^4$ scaling, is to model the localized
 %       excitation by the anharmonic oscillators of different
 %       frequency~\cite{PhysRevB.67.094203,ji2018theory}. In this
 %       approach, the interactions between the oscillators are essential
 %       to reproduce the $\omega^4$ scaling. This is seemingly very
 %       different from our theory where the interactions do not appear
 %       explicitly.  Most optimistically, with a proper
 %       ``renomalization'' of the interactions, the oscillator model may
 %       be mapped to a non-interacting model that has the similar form of
 %       the density of state of the mean-field model,
 %       eq.~(\ref{164344_2Aug18}), which may justify our assumption. This
 %       would be addressed in the future work.

\end{itemize}

% In the future work, one should address $\varepsilon_{\rm ex0}$.  This
% may be done using the formalism of the

% In summary, we empirically introduce the effects of the localization
% into the mean calculation, which correctly reproduces the numerical
% results of $D(\omega)$ in three dimension.

% For $D(\omega)$, we get $D(\omega)\sim \PP(0)\delta
% z^{-4}\omega^4$. This is consistent with the recent numerical
% observation~\cite{shimada2018spatial}. The pre-factor $\PP(0)$ may
% depend on the preparation protocol and interaction potential. In
% general, we get the smaller $\PP(0)$ for the slower quench
% rate~\cite{wang2018low}.

% If $\PP(\varepsilon)$ is not finite at $\varepsilon=0$, the above result should be
% modified.  For instance when $\PP(\varepsilon) =A\varepsilon^{-\alpha}$, we get
% $\rho(\lambda)\sim Ap_*^{\alpha-2}\lambda^{\frac{3}{2}-\alpha}$ or
% $D(\omega)\sim Ap_*^{\alpha-2}\omega^{4-2\alpha}$. In particular, when
% $\alpha=1/2$, we get $D(\omega)\sim \omega^3$, which may correspond to
% the random pinning~\cite{angelani2018probing}.

\begin{acknowledgments}
 We thank F.~Zamponi, P.~Urbani, A.~Ikeda and
 H.~Mizuno for kind discussions. This project has received funding from
 the European Research Coineduncil (ERC) under the European Union's
 Horizon 2020 research and innovation programme (grant agreement
 n°723955-GlassUniversality). 
\end{acknowledgments}

\bibliography{apssamp}
\end{document}